# The Revolution in Database Architecture

Jim Gray

Microsoft Research

Microsoft Research

Microsoft Corporation

One Microsoft Way

Redmond, WA  98052

# The Revolution in Database Architecture

Jim Gray
Microsoft
455 Market St. #1650
San Francisco, CA, 94105 USA
http://research.microsoft.com/~Gray
Gray@Microsoft.com

## ABSTRACT

Database system architectures are undergoing revolutionary changes. Most importantly, algorithms and data are being unified by integrating programming languages with the database system. This gives an extensible object-relational system where non-procedural relational operators manipulate object sets. Coupled with this, each DBMS is now a web service. This has huge implications for how we structure applications. DBMSs are now object containers. Queues are the first objects to be added. These queues are the basis for transaction processing and workflow applications. Future workflow systems are likely to be built on this core. Data cubes and online analytic processing are now baked into most DBMSs. Beyond that, DBMSs have a framework for data mining and machine learning algorithms. Decision trees, Bayes nets, clustering, and time series analysis are built in; new algorithms can be added. There is a rebirth of column stores for sparse tables and to optimize bandwidth. Text, temporal, and spatial data access methods, along with their probabilistic reasoning have been added to database systems. Allowing approximate and probabilistic answers is essential for many applications. Many believe that XML and xQuery will be the main data structure and access pattern. Database systems must accommodate that perspective. External data increasingly arrives as streams to be compared to historical data; so stream-processing operators are being added to the DBMS. Publish-subscribe systems invert the data-query ratios; incoming data is compared against millions of queries rather than queries searching millions of records. Meanwhile, disk and memory capacities are growing much faster than their bandwidth and latency, so the database systems increasingly use huge main memories and sequential disk access. These changes mandate a much more dynamic query optimization strategy – one that adapts to current conditions and selectivities rather than having a static plan. Intelligence is moving to the periphery of the network. Each disk and each sensor will be a competent database machine. Relational algebra is a convenient way to program these systems. Database systems are now expected to be self-managing, self-healing, and always-up. We researchers and developers have our work cut out for us in delivering all these features.

## 1. INTRODUCTION

This is an extended abstract for a SIGMOD 2004 keynote address. It argues that databases are emerging from a period of relative stasis where the agenda was "implement SQL better." Now database architectures are in the punctuated stage of punctuated-equilibrium. They have become the vehicles to deliver an integrated application development environment, to be data-rich nodes of the Internet, to do data discovery, and to be self-managing. They are also our main hope to deal with the information avalanche hitting individuals, organizations, and all aspects of human organization. It is an exciting time! There are many exciting new research problems and many challenging implementation problems. This talk highlights some of them.

## 2. THE REVOLUTIONS
### 2.1 Object Relational Arrives

We be data. But, you cannot separate data and algorithms. Unfortunately, Cobol has a data division and a procedure division and so it had separate committees to define each one. The database community inherited that artificial division from the Cobol Data Base Task Group (DBTG). We were separated from our procedural twin at birth. We have been trying to reunite with it for 40 years now. In the mid-eighties stored procedures were added to SQL (thank you Sybase), and there was a proliferation of object-relational database systems. In the mid-nineties many SQL vendors added objects to their own systems. Although these were each good efforts, they were fundamentally flawed because de novo language designs are very high risk.

The object-oriented language community has been refining its ideas since Simula67. There are now several good OO languages with excellent implementations and development environments (Java and C# for example.) There is a common language runtime that supports nearly all languages with good performance.

The big news now is the marriage of databases and these languages. The runtimes are being added to the database engine so that now one can write database stored-procedures (modules) in these languages and can define database objects as classes in these languages. Database data can be encapsulated in classes and the language development environment allows you to program and debug SQL seamlessly mixing Java or C# with SQL, doing version control on the programs, and generally providing a very productive programming environment. SQLJ is a very nice integration of SQL and Java, but there are even better ideas in the pipeline.

This integration of languages with databases eliminates the inside-the-database outside-the-database dichotomy that we have lived with for the last 40 years. Now fields are objects (values or references); records are vectors of objects (fields); and tables are se-quences of record objects. Databases are collections of tables. This objectified view of database systems has huge leverage – it enables most of the other revolutions. It is a way for us to structure and modularize our systems.

A clean object-oriented programming model also makes database triggers much more powerful and much easier to construct and debug. Triggers are the database equivalent of rule-based programming. As such, they have proponents and opponents. Having a good language foundation will probably not sway the active database opponents, but it will certainly make it easier to build systems.

The database integration with language runtimes is only possible because database system architecture has been modularized and rationalized. This modularity enables the other architectural revolutions which are done as extensions to the core data manger.

## 2.2  Databases are Web Services --TPlite

Databases are encapsulated by business logic. Before the advent of stored-procedures, all the business logic ran in the transaction processing monitor which was the middle tier of the classic three-tier presentation-application-data architecture. With stored procedures, the TP-monitors were disintermediated by two-tiered client/server architectures. The emergence of web-servers and HTTP brought three-tier architectures back to center stage – in part as protocol converters between HTTP and the database client/server protocol, and in part by moving the presentation services (HTML) back to the web server.

As eCommerce evolves, most web clients are application programs rather than browsers blindly displaying whatever the server delivers. Today, most eCommerce clients *screen-scrape* to get data from the web pages, but there is increasing use of XML web services as a way of delivering data to fat-client applications. Most web services are being delivered by classic web servers today (Apache, Microsoft IIS); but, database systems are starting to listen to port 80 and to publish web services. In this new world, one can take a class or a stored procedure implemented inside the database system, and publish it on the internet as a web service (WSDL interface definition, DISCO discovery, UDDI registration, and SOAP call stubs are all generated automatically). So, the TPlite client-server model is back, if you want it.

Designers still have the option of three-tier or n-tier application designs; but, they now have the two-tier option again. The simplicity of two-tier client/server is attractive, but security issues (databases have huge attack surfaces) may cause many designers to want three-tier server architectures with the web server in the demilitarized zone (DMZ).

It is likely that web services will be the way we federate heterogeneous database systems. This is an active research area. What is the right object model for a database? What is the right way to represent information on the wire? How do schemas work in the Internet? How does schema evolution work? How do you find data and databases? We do not have good answers to any of these questions. Much of my time is devoted to trying to answer these questions for the federation of astronomy databases we call the World-Wide Telescope.

## 2.3  Queues, Transactions, Workflows

The Internet is a loosely coupled federation of computer servers and clients. Clients are sometime disconnected, and yet they need to be able continue functioning. Rather than building tightly-coupled RPC-based applications, Internet-scale applications must be constructed as asynchronous tasks structured as workflows involving multiple autonomous agents. eMail gives an intuitive understanding of these design issues. You want to be able to read and send mail even though you are not connected to the network.

All the major database systems now include a queuing system that makes it easy to define queues, queue and dequeue messages, attach triggers to queues, and dispatch tasks driven by the queues. A good programming environment within the database system and the simplicity of the transaction model makes it easy and natural to use queues. Being able to publish queues as web services is also a big advantage. But, queues are almost immediately used to go beyond simple ACID transactions and implement publish-subscribe and workflow systems. These are built as applications atop the basic queuing system. There is a lot of innovation and controversy over exactly how workflows and notifications should work – it is an area of ferment and fruitful experimentation.

The research question here is how to structure workflows. Frankly, solutions to this problem have eluded us for several decades. But the immediacy of the problem is likely to create enough systems that some design patterns will emerge. The research challenge is to characterize these design patterns.

## 2.4  Cubes and Online Analytic Processing

Early relational systems used indices as table replicas that allowed vertical partitioning, allowed associative search, and allowed convenient data ordering. Database optimizers and executors use semi-join on these structures to run common queries on covering indices. These query strategies give huge speedups.

These early ideas evolved to materialized views (often maintained by triggers) that went far beyond simple covering indices and provided fast access to star and snowflake schema. In the 1990s we discovered the fairly common OLAP pattern of data cubes that aggregate data along many dimensions. The research community extended the cube-dimension concepts and developed algorithms to automate cube design and implementation. There are very elegant and efficient ways to maintain cubes. Useable cubes that aggregate multi-terabyte fact tables can be represented in a few gigabytes. These algorithms are now key parts of the major database engines. This is an area intense research and rapid innovation – much of the work now focuses on better ways to query and visualize cubes.

## 2.5  Data Mining

We are slowly climbing the value chain from data to information to knowledge to wisdom. Data mining is our first step into the knowledge domain. The database community has found a very elegant way to embrace and extend machine learning technology like clustering, decision trees, Bayes nets, neural nets, time series analysis, etc... The key idea is to create a *learning table* T; telling the system to learn columns $x, y, z,$ from attributes $a, b, c$ (or to cluster attributes $a, b, c,$ or to treat $a$ as the time stamp for $b$.) Then one inserts training data into the table T, and the data mining algorithm builds a decision tree or Bayes net or time series model for the data. The training phase uses SQL's well under-stood Create/Insert metaphor. At any point, one can ask the system to display the model as an XML document that, in turn, can be rendered in intuitive graphical formats.

After the training phase, the table T can be used to generate synthetic data; given a key $a,b,c$ it can return the likely $x,y,z$ values of that key along with the probabilities. Equivalently, T can evaluate the probability that some value is correct. The neat thing about this is that the framework allows you to add your own machine-learning algorithms to this framework. This gives the machine-learning community a vehicle to make their technology accessible to a broad user base.

Given this framework, the research challenges are now to develop better mining algorithms. There is also the related problem of probabilistic and approximate answers that is elaborated later.

## 2.6 Column Stores

It is increasingly common to find tables with thousands of columns – they arise when a particular object has thousands of measured attributes. Not infrequently, many of the values are null. For example, an LDAP object has 7 required and a thousand optional attributes. It is convenient to think of each object as a row of a table, but representing it that way is very inefficient – both in space and bandwidth. Classical relational systems represent each row as a vector of values and often materialize rows even if they are null (not all systems do that, but most do.) This row-store representation makes for very large tables and very sparse information.

Storing sparse data column-wise as ternary relations (key, attribute, value) allows extraordinary compression—often as a bitmap. Querying such bitmaps can reduce query times by orders of magnitude – and enable whole new optimization strategies. Adabase and Model204 pioneered these ideas, but they are now having a rebirth. The research challenge is to develop automatic algorithms that do column store physical design and to develop efficient algorithms for updating and searching column stores.

## 2.7 Text, Temporal, and Spatial Data Access

The database community has insulated itself from the information retrieval community, and has largely eschewed dealing with messy data types like time and space (not everyone has, just most of us.) We had our hands full dealing with the "simple stuff" of numbers, strings, and relational operators on them. But, real applications have massive amounts of text data, have temporal properties, and have spatial properties.

The DBMS extensibility offered by integrating languages with the DBMS makes it relatively easy to add data types and libraries for text, spatial, and temporal indexing and access. Indeed the SQL standard has been extended in all these areas. But, all three of these data types, and especially text retrieval, require that the database deal with approximate answers and with probabilistic reasoning. This has been a stretch for traditional database systems. It is fair to say that much more research is needed to seamlessly integrate these important data types with our current frameworks. Both data mining and these complex datatypes depend on approximate reasoning – but we do not have a clear algebra for it.

## 2.8 Semi-Structured Data

Not all data fits into the relational model. Jennifer Widom observes that we all start with the schema `<stuff/>` and then add structure and constraints. Even the best designed database leaves out some constraints and leaves some relationships unspecified.

A huge battle is raging in the database community. The radicals believe cyberspace is just one big XML document that should be manipulated with xQuery++. The reactionaries believe that structure is your friend and that semi-structured data is a mess to be avoided. Both camps are well represented within the database community – often stratified by age. It is easy to say that the truth lies somewhere in between, but it is hard at this point to say how this movie will end.

One especially interesting development is the integration of database systems with file systems. Individuals have hundreds of thousands of files (mails, documents, photos, ...). Corporations have billions of files. Folder hierarchies and traditional filing systems are inadequate – you just can't find things by location (folder) or grep (string search). A fully indexed semi-structured database of the objects is needed to for decent precision and recall on search. It is paradoxical, but file systems are evolving into database systems. These modern file systems are a good example of the semi-structured data challenge, and indeed are challenging some of the best data management architects.

## 2.9 Stream Processing

Data is increasingly generated by instruments that monitor the environment – telescopes looking at the heavens, DNA sequencers decoding molecules, bar-code readers watching passing freight-cars, patient monitors watching the life-signs of a person in the emergency room, cell-phone and credit-card systems looking for fraud, RFID scanners watching products flow through the supply chain, and smart-dust sensing its environment.

In each of these cases, one wants to compare the incoming data with the history of an object. The data structures, query operators, and execution environments for such stream processing systems are qualitatively different from classic DBMS architectures. In essence, the arriving data items each represent a fairly complex query against the existing database. Researchers have been building stream processing systems, and their stream-processing ideas have started appearing in mainstream products.

## 2.10 Publish-Subscribe and Replication

Enterprise database architects have adopted a wholesale-retail data model where data-warehouses collect vast data archives and publish subsets to many data-marts each of which serves some special interest group. This bulk publish-distribute-subscribe model is widely used and employs just about every replication scheme you can imagine. There is a trend to install custom subscriptions at the warehouse – application designers are adding thousands, sometimes millions of subscriptions. In addition, they are asking that the subscriptions have real-time notification. That is, when new data arrives, if it affects the subscription, then the change is immediately propagated to the subscriber. For example, finance applications want to be notified of price fluctuations, inventory applications want to be notified of stock level changes, and information retrieval applications want to be notified when new content is posted.

Pub-sub and stream processing systems have similar structure. The millions of standing queries are compiled into a dataflow graph. As new data arrives, the data flow graph is incrementally evaluated to see which subscriptions are affected. The new data triggers updates to those subscriptions. This technology relies heavily on the active-database work of the 1990s and is still evolving. The research challenge is to support more sophisticated standing queries and to provide better optimization techniques that handle the vast number of queries and vast data volumes.

## 2.11  Late Binding in Query Plans

All these changes have a huge impact on the way the database query optimizer works. Having user-defined functions deep inside the query plans makes cost estimation problematic. Having real data with high skew has always been problematic, but in this new world the relational operators are just the outer loop of a non-procedural program that should be executed with the least cost and in parallel.

Cost-based static-plan optimizers continue to be the mainstay for simple queries that run in seconds. But, for complex queries, the query optimizer must adapt to current workloads, must adapt to data skew and statistics, and must plan in a much more dynamic way – changing plans as the system load and data statistics change. For petabyte-scale databases it seems the only solution is to run continuous data scans and let queries piggyback on the scans. Teradata pioneered that mechanism, and it is likely to become more common in the future.

## 2.12  Massive Memory, Massive Latency

To make life even more interesting, disk and memory capacities continue to grow faster than latency and bandwidth improve. It used to take less than a second to read all of ram and less than 20 minutes to read everything on a disk. Now, a multi-terabyte ram memory scans take minutes and terabyte-disk scans take hours. Random access is a hundred times slower than sequential. These changing ratios require new algorithms that intelligently use multi-processors sharing a massive main memory, and intelligently use precious disk bandwidth. The database engines need to overhaul their algorithms to deal with the fact that main memories are huge (billions of pages trillions of bytes). The era of main-memory databases has finally arrived.

## 2.13  Smart Objects: Databases Everywhere

At the other extreme, each disk controller now has tens of megabytes of storage and a very capable processor. It is quite feasible to have intelligent disks that offer either database access (SQL or some other non-procedural language) and even web service access. Moving from a block-oriented disk interface to a file interface, and then to a set or service interface has been the goal of database machine advocates for three decades. In the past they needed special purpose hardware. But, now disks have fast general purpose processors as a consequence of Moore's law. So, it seems likely that database machines will have a rebirth.

In a related development, people building sensor networks have discovered that if you view each sensor as a row of a table, where the sensor values are fields of the row, then it is very easy to write programs to query the sensors. What's more, distributed query technology, augmented with some new algorithms gives very efficient programs for these sensor networks, minimizing bandwidth and making them easy to program and debug. So tiny-database systems are appearing in smart dust – a surprising and exciting development.

## 2.14  Self Managing and Always Up

If every file system, every disk and every piece of smart dust has a database inside, database systems will have to be self-managing, self-organizing, and self healing. The database community is rightly proud of the advances they have made in automating design and operation – most people are unaware that their eMail system is a simple database and that their file system is a simple database and that many other applications they use and manage are in fact simple database systems. But, as you can see from the feature list enumerated here, database systems are becoming much more sophisticated. Much work remains to make the distributed data stores so robust that they never lose data and they always answer questions efficiently.

## 3.  CONCLUDING REMARKS

The theme of this talk is that we live in a time of extreme change. It is an exciting time; essentially everything all design assumptions are being re-evaluated. There are research challenges everywhere. There are no small challenges in this list of revolutions. Yet, I think our biggest challenge is a unification of approximate and exact reasoning. Most of us come from the exact-reasoning world—but most of our clients are asking questions with approximate or probabilistic answers.

The restructuring of database systems to be web services and to integrate with language runtimes has created a modularity that enables these revolutions. The reunification of code and data is pivotal. Almost all the other changes depend on that. The extension framework allows researchers and entrepreneurs to add new algorithms and whole new subsystems to the DBMS. Databases are evolving from SQL-engines to data integrators and mediators that provide a transactional and non-procedural access to data in many forms. Database systems are becoming database operating systems, into which one can plug subsystems and applications.

The database community has a healthy interplay between research and development. Virtually all the people and most innovations in database systems can be traced to the research prototypes first describe in research papers. Product groups watch research prototypes with great interest, academics frequently take sabbaticals in industry, and there are many startups. These collaborations are world-wide, largely fostered by SIGMOD and the VLDB-Foundation's international focus. The ecosystem compensates for the haphazard government funding of database research. Data and databases are central to all aspects of science and industry – and researchers and industry recognizes that, even if funding agencies do not.

Going forward, the information avalanche shows no sign of slowing. This guarantees a full menu of challenges for the database research community -- challenges far beyond the ones mentioned here. But, I believe the low-hanging fruit is clustered around the topics outlined here.

## 4.  ACKNOWLEDGMENTS